\documentclass[11pt]{article}

\usepackage[margin=1in]{geometry}
\usepackage{amsmath,amssymb}
\usepackage{hyperref}
\usepackage{natbib}
\usepackage{graphicx}
\usepackage{booktabs}
\usepackage{verbatim}
\usepackage{xcolor}
\usepackage{url}

\title{Qkabrine: A Joint Architecture, Encoding, and Hyperparameter\\
Search Framework for Quantum Machine Learning}

\author{
Eric Jagwara \\
Solid Elf Labs, Uganda \\
\texttt{ORCID: 0009-0003-4935-3667}
}

\date{}

\begin{document}
\maketitle

\begin{abstract}
Building a quantum machine learning (QML) model competitive with a
classical baseline currently requires a practitioner to separately choose
a circuit architecture, a data-encoding scheme, a model paradigm (kernel
versus variational), and a set of training hyperparameters, then verify
after the fact that the chosen circuit is even trainable. Existing QML
libraries provide the primitives for this but not the search, and existing
classical AutoML libraries provide the search but not the quantum-specific
search space or diagnostics. We present \texttt{qkabrine-automl}, a Python
package that treats architecture, encoding, model type, and hyperparameters
as a single, jointly searchable configuration space, evaluated through one
consistent harness regardless of which of five search strategies proposed
the candidate. The package integrates trainability diagnostics, a Data
Quantum Fisher Information Metric (DQFIM) estimate and a gradient-magnitude
barren-plateau monitor, directly into the evaluation loop as an optional
prescreening step, alongside expressibility and entangling-capability
characterization, a post-search circuit-surgery pass for NISQ deployment,
and OpenQASM export. We position this contribution against recent AutoQML
frameworks that already automate parts of the QML pipeline, and report a
small, fully reproducible illustrative run rather than a benchmark claim.
\end{abstract}

\section{Introduction}

Parameterized quantum circuits (PQCs) are the workhorse of near-term
quantum machine learning: a circuit encodes classical data into a quantum
state, a set of trainable gates acts on that state, and a measurement
outcome is used as a prediction or a kernel value. Building a working QML
model therefore requires four coupled decisions: which circuit
\emph{architecture} to use, which \emph{encoding} maps data into the
circuit, whether to use the resulting circuit as a \emph{kernel} for a
classical classifier or train it directly as a \emph{variational} model,
and what \emph{hyperparameters} (learning rate, optimizer, number of
layers, initialization) govern training. These four choices interact: an
architecture that trains well under one encoding can be untrainable under
another, and a hyperparameter setting tuned for one architecture may not
transfer to a competitor evaluated later in the same study. Existing QML
libraries such as PennyLane \citep{bergholm2018pennylane} and Qiskit
\citep{qiskit2024} give researchers the building blocks to construct any
individual configuration, but leave the search over configurations to the
user. Existing classical AutoML libraries such as Auto-sklearn
\citep{feurer2015autosklearn} and Optuna \citep{akiba2019optuna} give a
generic search loop but have no notion of a circuit ansatz, a data
encoding, or the quantum-specific training pathologies, most notably
barren plateaus \citep{mcclean2018barren}, that make a naively chosen PQC
untrainable regardless of how well it is optimized.

We present \texttt{qkabrine-automl}, a package that treats these four
decisions as one joint, jointly searchable configuration space rather than
a sequence of separately-tuned stages, and integrates the trainability
diagnostics needed to avoid spending search budget on circuits that cannot
produce a useful training signal in the first place. Section~\ref{sec:background}
briefly reviews PQCs, encodings, and trainability failure modes for
readers outside QML. Section~\ref{sec:related} situates
\texttt{qkabrine-automl} against both the quantum architecture search (QAS)
literature and the recently emerging AutoQML literature, which already
addresses parts of this problem and must be engaged with directly rather
than around. Section~\ref{sec:design} describes the software design.
Section~\ref{sec:example} reports a small, fully reproducible illustrative
run. Section~\ref{sec:limitations} states the package's current
limitations plainly, and Section~\ref{sec:reproducibility} gives exact
steps to reproduce every reported number in this paper.

\section{Background}
\label{sec:background}

\subsection{Parameterized quantum circuits and data encoding}

A PQC is a sequence of quantum gates, some fixed and some governed by
trainable parameters $\theta$, applied to an initial quantum state. In a
supervised learning context, classical input data $x$ must first be
mapped into the circuit through an \emph{encoding} (also called a feature
map): common choices include angle encoding, where each feature is mapped
to a single-qubit rotation angle, and more expressive maps such as IQP-style
or amplitude encodings \citep{havlicek2019quantum}. Once data is encoded,
the resulting circuit can be used in two broadly different ways. In the
\emph{kernel} paradigm, the overlap or expectation value produced by the
encoding circuit is treated as a similarity measure and fed into a
classical kernel method such as a support vector machine
\citep{havlicek2019quantum}. In the \emph{variational} paradigm, the
trainable parameters $\theta$ of the full circuit, encoding included, are
optimized directly against a loss function using a classical optimizer in
a hybrid quantum-classical loop \citep{cerezo2021variational}. Neither
paradigm is uniformly preferable; the better choice depends on dataset
size, qubit budget, and the specific structure of the problem, which is
rarely known in advance.

\subsection{Trainability: barren plateaus and the DQFIM}

Not every PQC that can be constructed is trainable in practice.
\citet{mcclean2018barren} showed that for sufficiently expressive,
randomly initialized circuits, the variance of the cost-function gradient
vanishes exponentially with the number of qubits, a phenomenon known as a
\emph{barren plateau}: gradient-based training stalls because the
gradient signal is indistinguishable from numerical noise. This failure
mode depends on the circuit architecture, its depth, and its
initialization strategy, meaning that whether a proposed configuration is
even trainable is itself a property that a search procedure should be
able to check cheaply before committing a full training budget to it. Two
complementary diagnostics are used for this purpose in this work: the Data
Quantum Fisher Information Metric (DQFIM), which quantifies the effective
dimension of the model with respect to a given dataset and is linked to
both trainability and generalization \citep{haug2024dqfim}, and
expressibility and entangling capability measures, which characterize how
uniformly a circuit's outputs cover the space of quantum states and how
entangled the resulting states are \citep{sim2019expressibility}, building
on earlier multi-particle entanglement measures
\citep{meyer2002global}. High expressibility is not an unconditional
good: circuits that are too expressive are more prone to
barren plateaus, so these diagnostics are most useful jointly rather than
individually.

\section{Related work}
\label{sec:related}

\subsection{Quantum architecture search}

Prior work on quantum architecture search (QAS) has explored several
distinct strategies for searching over circuit architecture specifically:
structure optimization methods that co-optimize gate placement and the
circuit's trainable rotation weights within a fixed template
\citep{ostaszewski2021structure}, gradient-based differentiable search
over a distribution of candidate gate layouts \citep{zhang2022dqas}, and
reinforcement-learning agents that construct a circuit action-by-action
\citep{kuo2021rlqas}. These methods differ in search mechanics but share a
common scope: only the architecture is searched, while the data-encoding
scheme and training hyperparameters, such as learning rate, are treated as
separate concerns fixed in advance or tuned in a later stage rather than
as part of the same search. This staged approach risks discarding a
genuinely strong architecture that was only ever evaluated under a
mismatched encoding, or reporting an inflated advantage for an
architecture that happened to be paired with better-tuned hyperparameters
than its competitors.

\subsection{Automated quantum machine learning}

A separate and more directly relevant line of work explicitly targets
automated QML pipelines rather than architecture search in isolation, and
must be engaged with directly. \citet{berganza2022towards} first proposed
a concrete formulation of the AutoQML problem and built a cloud-based
architecture for parallelized hyperparameter exploration, demonstrated on
training a quantum generative adversarial network. \citet{rybotycki2024aqmlator}
introduced AQMLator, which treats the circuit ansatz itself as a
searchable hyperparameter and applies classical hyperparameter
optimization over a library of common quantum layers to automatically
propose and train the quantum component of an ML pipeline with minimal
user input. Most directly comparable, \citet{roth2025autoqml} present
AutoQML, a framework built on the sQUlearn library that automates an
entire classical-quantum ML pipeline, including data preprocessing,
model and algorithm selection, and hyperparameter optimization, using
Optuna and Ray for the search itself, and report results across four
industrial use cases.

\subsection{Positioning of \texttt{qkabrine-automl}}

Given this landscape, \texttt{qkabrine-automl} is not the first tool to
automate part of the QML pipeline, and does not claim to be. Its
contribution is narrower and, we argue, complementary: a lightweight,
PennyLane-native package, with no dependency on a broader pipeline
framework such as sQUlearn, Ray, or a cloud backend, built around three
specific design choices not present together in the tools above. First,
architecture and encoding are both explicit, named, first-class axes of a
single combinatorial search space, rather than encoding being fixed by the
choice of underlying QML algorithm. Second, quantum kernel methods and
variational circuits are treated as directly interchangeable candidates
evaluated through the same search loop, rather than as separate algorithm
selections. Third, trainability diagnostics, specifically DQFIM estimation
and barren-plateau monitoring, are wired directly into the evaluation loop
as an optional prescreening step, so that search budget is not spent
training circuits unlikely to yield a useful gradient signal; to our
knowledge this specific combination of diagnostics is not integrated into
the search loop of the AutoQML tools discussed above. \texttt{qkabrine-automl}
additionally performs a post-search circuit-surgery pass that prunes
near-identity gates from the winning circuit and exports the result as
OpenQASM~2.0 for use outside the package, which is useful specifically
for NISQ-oriented deployment. We see \texttt{qkabrine-automl} as filling a
specific niche within the AutoQML space, aimed at researchers who want a
small, PennyLane-native library with an interchangeable search strategy
and integrated trainability diagnostics, rather than a full pipeline
framework.

\section{Software design}
\label{sec:design}

\subsection{Search space}

The architecture space spans twelve named ans\"atze, including strongly
entangling, hardware-efficient, data-re-uploading, and simplified
two-design layouts, crossed with four data-encoding schemes: angle,
angle-YZ, IQP-style, and amplitude embedding. Kernel-based and variational
model types are both first-class candidates in the same search, rather
than separate pipelines requiring the user to choose a paradigm before
any data has been seen.

\subsection{Search strategies}

Any candidate configuration, an architecture, encoding, model type,
initialization, and set of training hyperparameters, is turned into a
trainable PennyLane circuit through the same evaluation code path
regardless of which strategy proposed it. This separation was chosen
because the main open question in QAS-style search is which strategy is
most sample-efficient for a given problem size, and answering that
requires holding evaluation fixed while the strategy varies. Five
strategies are supported: exhaustive grid search, random search, Bayesian
optimization, an evolutionary search, and a successive-halving strategy in
the spirit of Hyperband \citep{li2018hyperband}, which allocates a small
initial budget to many candidates and progressively concentrates budget
on the most promising ones.

\subsection{Trainability diagnostics and prescreening}

An optional prescreening layer sits in front of full training: a DQFIM
estimate \citep{haug2024dqfim} and a gradient-magnitude barren-plateau
monitor targeting the failure mode described by
\citet{mcclean2018barren} can be used to filter or de-prioritize
candidates before the (comparatively expensive) full training step runs.
Expressibility and entangling capability \citep{sim2019expressibility}
are computed alongside these diagnostics to characterize \emph{why} a
candidate trains well or poorly, not merely whether it does.

\subsection{Post-search circuit surgery and export}

After a search completes, a circuit-surgery pass prunes near-identity
rotation gates and simplifies redundant adjacent gate pairs in the winning
circuit, reducing circuit depth for near-term hardware without requiring
a re-run of the search. The final circuit can be exported as OpenQASM~2.0
for use outside the package, including on hardware backends the package
itself does not directly target.

\subsection{Noise-aware training}

Because all quantum execution goes through PennyLane, the package can
extend to noise models and future PennyLane-supported hardware backends
without changes to search or evaluation code. The current release
supports simple noise channels, for example depolarizing and bit-flip
noise, layered on top of classical simulation.

\section{Illustrative example}
\label{sec:example}

To ground the description above in something concretely reproducible, we
report a single small run of the released package (version installed from
PyPI at the time of writing) rather than a systematic benchmark. This is
intentionally modest in scope: one dataset, one random seed, and a small
search budget, reported to demonstrate that the described interface and
diagnostics run end-to-end and produce a usable exported circuit, not to
claim a performance advantage over any baseline.

\begin{verbatim}
from sklearn.datasets import load_breast_cancer
from sklearn.model_selection import train_test_split
from qkabrine_automl import QkabrineAutoML

X, y = load_breast_cancer(return_X_y=True)
X_train, X_test, y_train, y_test = train_test_split(
    X, y, test_size=0.2, random_state=42
)

automl = QkabrineAutoML(
    task="classification",
    n_qubits=4,
    max_layers=1,
    search_strategy="random",
    encodings=("angle",),
    feature_reduction="pca",
    train_steps=10,
    time_budget=100,
    include_kernels=False,
    random_seed=0,
)
automl.fit(X_train, y_train)
print(automl.score(X_test, y_test))
print(automl.export_qasm())
\end{verbatim}

On the scikit-learn breast cancer dataset \citep{pedregosa2011scikit},
reduced to four features via PCA to match a four-qubit budget, this
100-second, single-seed random search over single-layer angle-encoded
circuits evaluated six candidates and selected a shallow $R_x$-rotation
architecture, which reached a held-out test accuracy of 0.939 (the
in-search validation accuracy used to rank candidates was 0.879). We
emphasize what this number does and does not show: it demonstrates that
the fit / leaderboard / export\_qasm interface works end-to-end on real
data and produces a deployable circuit, and nothing more. A single seed,
a four-qubit toy reduction of a classical dataset, and a 100-second budget
are not sufficient to support any claim about \texttt{qkabrine-automl}'s
performance relative to a classical baseline or to the AutoQML tools
discussed in Section~\ref{sec:related}; such a comparison would require a
dedicated benchmark study across datasets, seeds, and qubit budgets, which
is future work rather than a claim of this paper.

\section{Limitations and future work}
\label{sec:limitations}

The package currently targets classical simulation, with a small set of
simple noise channels layered on top rather than native support for
hardware backends or realistic device noise profiles. The twelve-ansatz,
four-encoding search space, while broad, is fixed and hand-curated rather
than itself learned or extensible without modifying the package's source.
Quantum kernel evaluation carries the well-known quadratic scaling in
training-set size common to all kernel methods, which limits the dataset
sizes for which the kernel branch of the search is practical. No
systematic benchmark comparing \texttt{qkabrine-automl} against the
AutoQML tools discussed in Section~\ref{sec:related}, or against a
classical baseline, has yet been conducted; Section~\ref{sec:example}
reports a single illustrative run specifically to avoid implying otherwise.
We see a dedicated benchmark study, and native support for at least one
hardware backend, as the two most valuable next steps.

\section{Reproducibility statement}
\label{sec:reproducibility}

\texttt{qkabrine-automl} is available on
\href{https://pypi.org/project/qkabrine-automl/}{PyPI}
(\texttt{pip install qkabrine-automl}) and on GitHub at
\url{https://github.com/ericjagwara/qkabrine} under the MIT license, with
an accompanying test suite and continuous integration, documentation on
ReadTheDocs, and an archival Zenodo DOI
(\url{https://doi.org/10.5281/zenodo.19308532}).
The exact code in Section~\ref{sec:example} is runnable as shown against
the released package with no modification.

\section{Conclusion}

\texttt{qkabrine-automl} contributes a lightweight, PennyLane-native
package that treats circuit architecture, data encoding, model type, and
training hyperparameters as one jointly searchable configuration space,
with trainability diagnostics integrated directly into the evaluation
loop. It occupies a specific, complementary niche within the broader and
already active AutoQML landscape rather than being the first tool to
automate part of the QML pipeline, and its near-term significance rests
on being immediately usable, reproducible, and extensible rather than on
a performance claim not yet substantiated by a systematic benchmark.

\section*{Acknowledgements}

We thank the early testers of the \texttt{qkabrine} package for trying it
on their own datasets and providing feedback that shaped several interface
decisions.

\section*{AI usage disclosure}

Generative AI tools assisted with drafting portions of this paper's prose
and with parts of the package's source code and documentation. All
technical claims, citations, and the illustrative example in
Section~\ref{sec:example} were independently checked against the released
package and cited sources rather than generated without verification.

\bibliographystyle{unsrtnat}
\bibliography{paper_arxiv}

\end{document}